\documentstyle[aps,preprint]{revtex}

\newcommand{\bA}{\mbox{\boldmath$ A $}}
\newcommand{\bB}{\mbox{\boldmath$ B $}}
\newcommand{\bE}{\mbox{\boldmath$ E $}}
\newcommand{\bU}{\mbox{\boldmath$ U $}}
\newcommand{\bV}{\mbox{\boldmath$ V $}}
\newcommand{\bX}{\mbox{\boldmath$ X $}}
\newcommand{\ba}{\mbox{\boldmath$ a $}}
\newcommand{\bb}{\mbox{\boldmath$ b $}}
\newcommand{\be}{\mbox{\boldmath$ e $}}
\newcommand{\bk}{\mbox{\boldmath$ k $}}
\newcommand{\bp}{\mbox{\boldmath$ p $}}
\newcommand{\bu}{\mbox{\boldmath$ u $}}
\newcommand{\bv}{\mbox{\boldmath$ v $}}
\newcommand{\bx}{\mbox{\boldmath$ x $}}
\newcommand{\by}{\mbox{\boldmath$ y $}}
\newcommand{\bz}{\mbox{\boldmath$ z $}}
\newcommand{\bmu}{\mbox{\boldmath$ \mu $}}
\newcommand{\Natural}{{\sf I \hspace{-0.15em} N}}

\newcommand{\Journal}[5]{#5\ {\em #1}#2 {\bf #3}\ #4}
\newcommand{\Paper}[3]{#1 #3}
\newcommand{\Book}[7]{#1 #5 {\em #2\/} (#4: #3)#7}
\newcommand{\InBook}[9]{#1 #8 {\em #3} #4 (#6: #5) #9}
\newcommand{\Preprint}[5]{#1 #5 #3\ #4}
\newcommand{\prva}{\Journal{Phys.\ Rev.\ }{A}}
\newcommand{\prvl}{\Journal{Phys.\ Rev.\ Lett.}{}}
\newcommand{\And}{and }  
\newcommand{\IBEds}[1]{ed #1  }

\begin{document}
\draft
\title{Product rule for gauge invariant Weyl symbols and its
application to the semiclassical description of guiding center
motion}
\author{ M.\ M\"uller\footnote{email:mmueller@eckart.lbl.gov}}
\address{Department of Physics and Lawrence Berkeley Laboratory\\
University of California, Berkeley, California 94720\thanks{
Present address: M.\ M\"uller, Petrinistr.\ 7, 97080 W\"urzburg}}
\date{Revised version November 17, 1998}
\maketitle

\begin{abstract}
\par
We derive a product rule for gauge invariant Weyl symbols
which provides a generalization of the well-known Moyal
formula to the case of non-vanishing electromagnetic fields.
Applying our result to the guiding center problem we
expand the guiding center Hamiltonian into an asymptotic power
series with respect to both Planck's constant $\hbar$ and an
adiabaticity parameter already present in the classical theory.
This expansion is used to determine the influence of quantum
mechanical effects on guiding center motion.
\end{abstract}

\section*{1.\ Introduction}

\label{secI}

In many physical applications charged particles are exposed
to strong time-independent magnetic fields
$\bB(\bx)$ and additional
electrostatic potentials $\phi(\bx)$.
Important examples are the magnetic confinement of plasmas,
trapping of ions in accelerator facilities \cite{Sav97} as
well as the Quantum Hall effect \cite{Pra90}.
Classically, the motion in such field configurations may be
visualized as a fast rotation in the plane perpendicular to
the magnetic field (``gyration''), with the center of the
circular orbit moving slowly parallel to the magnetic field
lines and drifting very slowly across both electric and
magnetic field lines (``guiding center motion'').
The underlying assumption that clearly distinguishable time
scales of motion exist is known as ``guiding center
approximation'' or, more general, ``adiabatic approximation.''

To separate gyration and guiding center motion, different
kinds of perturbative calculations have been applied in
classical mechanics \cite{Alf50,Nor63,Ban67,Nor78,Lit83}.
Adiabatic invariants and equations of motion for the guiding
center may be derived in a systematic way from Hamiltonian
theory \cite{Gar59,Myn79,Lit79,Lit81a,Wey86,Wey87}. 
For a semiclassical description of guiding center
motion the method invented by Littlejohn \cite{Lit79,Lit81a}
using non-canonical, but gauge invariant phase space
coordinates turns out to be the best starting point.
There, like in all classical investigations of the guiding
center problem, a dimensionless expansion parameter $\epsilon$
is introduced by replacing the electric charge $q$ with
$q/\epsilon$ \cite{Nor63,Kru65}.
Physically, $\epsilon$ represents the ratio of the gyroradius
to the scale lengths of the external fields and is interpreted
as an adiabatic parameter.
Employing symplectic geometrical techniques, relations between
the guiding center (phase space) coordinates and the particle's
position and velocity are obtained which take the form
of asymptotic power series in $\epsilon$.
After writing down the Hamiltonian in terms of the guiding
center coordinates, its dependence on the rapidly oscillating
gyration angle is removed by means of averaging Lie
transforms.
The equations of motion resulting from the  guiding
center Hamiltonian confirm that the magnetic moment caused
by the gyration is an adiabatic invariant.

In low-temperature experiments the total energies of the
particles are of order of the lowest Landau levels in the
magnetic field $\bB(\bx)$.
Therefore quantum mechanical effects have to be taken into
account when deriving equations of motion for the guiding
center.
So far this has been done only in the special case of a
charged particle in the magnetic field outside of a
rectilinear current filament \cite{Muel95a,Muel96}.
To determine explicitly the quantum corrections to guiding
center motion in arbitrary field configurations a method is
needed which results in an expansion of the quantized guiding
center Hamiltonian into a formal power series in both the
(classical) parameter $\epsilon$ and Planck's constant
$\hbar$.
The quantum guiding center theory developed by Maraner in
two inspiring papers \cite{Mar96b,Mar97} uses only
the magnetic length $l_B=\sqrt{\hbar c/(q 
|\bB|)}$ as expansion parameter.
The power series expansion of the guiding center
Hamiltonian operator with respect to $l_B$ does not distinguish
between terms of adiabatic origin already present in classical
mechanics and quantum corrections caused by the
non-commutativity of the operator algebra.
For experimental purposes, however, it is very important to
know whether the classical picture is valid even at low
temperatures or whether quantum effects dominate guiding
center motion.

A first step to answer this question lies in the observation
that there is great formal resemblance between guiding center
motion and adiabatic motion of neutral spinning particles in
an inhomogeneous magnetic field \cite{Fel80}.
The latter has recently been studied in more detail
because it represents a standard example for the occurrence of
``geometrical'' forces in dynamical systems 
\cite{Sha89,Ber86,Aha92,Ber93a}.
A semiclassical investigation of this motion \cite{Lit93b}
involves a multicomponent version of the Weyl calculus
\cite{Lit91a,Lit91b,Wei93}.
It has the appealing feature that two different
expansion parameters are used: one, $\epsilon_a$,
connected with adiabaticity (i.e.\ the assumption that the
magnetic field does not change appreciably during a precession
period) and another, $\epsilon_s$, proportional to
$\hbar$, controlling the validity of the semiclassical
approximation.
In the diagonalized Hamiltonian, which describes orbital motion, 
the potential terms are expanded with respect to
both $\epsilon_a$ and $\epsilon_s$.
To achieve the same goal for the guiding center Hamiltonian,
the Wigner-Weyl formalism \cite{Wey27,Wig32} therefore
seems to provide the appropriate tools.

In general the Weyl transform of a quantum mechanical operator
is a uniquely determined phase space function which may be 
defined as follows \cite{Bal84,Osb95,Ber91}:
Starting from the fundamental operators
$\hat{\bx}$ and $\hat{\bp}$,
a particular, continuously indexed basis $\hat\Delta
(\bx,\bp)$
of the operator space is constructed.
(Here, as in the following, the hat denotes an operator.)
The representation of an arbitrary operator $\hat A$ as a linear
combination of the operators $\hat\Delta(\bx,
\bp)$ involves $c$-number coefficients
which are labeled by the continuous variables
$\bx$ and $\bp$.
They constitute a function $A_W(\bx,
\bp)$ on phase space which is denoted as the
Weyl symbol of the operator $\hat A$.
The relation between the symbol $C_W(\bx,
\bp)$ associated with an operator product $\hat C = \hat A 
\hat B$ and the symbols $A_W(\bx,
\bp)$ and $B_W(\bx,
\bp)$ of its factors is given by a nontrivial composition rule
known as Moyal formula \cite{Moy49}.

If a magnetic field $\bB
(\bx)\!=\!\nabla\!\!\times\!\!
\bA(\bx)$
is present, however, the Weyl correspondence should be
re-defined, because the gauge dependence of the canonical
momentum $\hat{\bp}$ causes the basic operators $\hat \Delta
(\bx,\bp)$ to be gauge dependent as well.
As shown in \cite{Str56a} this leads to the undesirable
consequence that the Weyl symbol of a gauge invariant operator
becomes gauge dependent and that, vice versa, the operator
corresponding to a gauge invariant phase space function is
itself in general not gauge independent.
The most natural way to include the principle of gauge invariance 
into the Weyl formalism is to replace the (gauge dependent)
canonical momentum $\hat{\bp}$
appearing in the definition of $\hat \Delta
(\bx,\bp)$
by the (gauge invariant) kinetic momentum
$\hat{\bk} = m \hat{\bv}
= \hat{\bp} - (q/c)
\,\hat{\!\bA}(\hat{\bx})$
\cite{Str56a,Bia77,Ser86}.
The coefficient function of an operator $\hat A$ 
(not to be confused with the vector potential) with respect
to the new set of basic operators
$\hat \Delta(\bx,\bk)$
will be denoted in the following as the gauge invariant Weyl
symbol $A_W(\bx,\bk)$ of $\hat A$.
One can show that  $A_W(\bx,\bk)$ is a gauge invariant
phase space function if and only if $\hat A$ is gauge
invariant (for more details cf.\ \cite{Ser86} and
section 2).
Obviously, the product rule for gauge invariant Weyl
symbols will be different from the usual Moyal formula.

After comparing ordinary and gauge invariant Weyl calculus in a
little more detail, we will explicitly derive the
gauge invariant generalization of Moyal's formula in the
next section and discuss its most important properties.
In section 3, the gauge invariant Weyl formalism will be
applied to separate the different time scales occurring in the
motion of charged particles in external electromagnetic fields
within a semiclassical framework.
As a result we will expand both guiding center coordinates and
the guiding center Hamiltonian into asymptotic power series
with respect to the adiabatic parameter $\epsilon$ and Planck's
constant $\hbar$.
Section 4 contains a summary of our results and conclusions
concerning the influence of quantum effects on guiding center
motion which can be derived from our expansion of the guiding
center Hamiltonian.
Finally we compare our results to the quantum mechanical
calculations of Maraner. 
In the appendix, the classical guiding center theory for the
motion of a charged particle in a magnetic field of constant
direction and an additional electrostatic field is summarized. 

\section*{2.\ Product rule for gauge invariant Weyl symbols}

\label{secII}

In order to set the stage for our computations, let us briefly
review some basic features of the ordinary Weyl transform valid
in a six-dimensional flat phase space in the absence of magnetic
fields \cite{Bal84,Osb95,Ber91,Lea68a,McD88}.
Starting with the set of generating Heisenberg operators
\begin{equation}
\label{Tdef1}
\hat T\!\left(\bu,\bv
\right)
\,\equiv\,
\exp\!\left[\,i\!\left(
\bu\!\cdot\!\hat{\bp} +
\bv\!\cdot\!\hat{\bx}
\right)
\right],
\end{equation}
we introduce a basis 
\begin{eqnarray}
\label{Deltadef1}
\hat\Delta\!\left({\bx},
{\bp}\right)
&\equiv&
\left(\frac{\hbar}{2\pi}\right)^{\!3}\!
\int\!d^3\!\bu\,d^3\!\bv
\exp\!\left[\,i\!\left(
\bu\!\cdot\!\bp +
\bv\!\cdot\!\bx
\right)\right]
\hat T\!\left(-\bu,-\bv
\right) \nonumber \\
&=&
\left(\frac{\hbar}{2\pi}\right)^{\!3}\!
\int\!d^3\!\bu\,d^3\!\bv
\exp\!\left\{i\left[
\bu\!\cdot\!\left(
\bp - \hat{\bp}\right) +
\bv\!\cdot\!\left(
\bx-\hat{\bx}\right)
\right]\right\}
\end{eqnarray}
of the operator space which is labeled by the
continuous classical variables
$\bx$ and $\bp$.
If an operator $\hat A$ is written as a linear
combination of the $\hat\Delta\!\left(
\bx,\bp\right)$,
\begin{equation}
\label{Weylexpansion1}
\hat A = 
\int \frac{d^3\!\bx\,d^3\!\bp}
{h^3}
\,\hat\Delta\!\left({\bx},
{\bp}\right)
A_W\!\left(\bx, \bp\right)\,,
\end{equation}
the uniquely determined coefficient function
\begin{equation}
\label{Weylsymboldef1}
A_W\!\left(\bx, \bp\right) =
\int\! d^3\bx' \,\langle
\bx' | \hat A 
\,\hat\Delta\!\left(\bx,
\bp\right)|
\bx'\rangle
\equiv \mbox{Tr}\!\left[ \hat A
\,\hat\Delta\!\left({\bx},
{\bp}\right)\right]
\end{equation}
is called the Weyl symbol associated with $\hat A$.
Note that equation (\ref{Weylsymboldef1}) is a direct
consequence of definition (\ref{Weylexpansion1}) and
\begin{equation}
\label{Deltaprod}
\mbox{Tr}\!\left[\hat\Delta\!\left({\bx},
{\bp}\right)\hat\Delta\!
\left(\bx',\bp'
\right)\right]= ( 2 \pi \hbar)^{\!3} \delta
(\bp-\bp')
\,\delta(\bx -\bx')\,.
\end{equation}
Here, as in the following, we leave aside
questions of convergence and the mathematical
problem of characterizing the class of operators for which
expansions like (\ref{Weylexpansion1}) exist.

From equations (\ref{Tdef1})--(\ref{Weylsymboldef1}) and the
duplication formula
\begin{equation}
\label{duplication1}
\hat T\!\left(\bu,\bv
\right)
\hat T\!\left(\bu',\bv'
\right)
= \exp\!\!\left[\,i\,\frac{\hbar}{2}\,
\left(\bu\!\cdot\!\bv'
-\bu'\!\cdot\!\bv\right)\right]
\hat T\!\left(\bu+\bu',
\bv+\bv'\right),
\end{equation}
one can immediately determine
the relation between the Weyl symbol $C_W
(\bx,\bp)$ 
of a product operator $\hat C= \hat A \hat B$ and the symbols
of its factors. 
The result is the well-known Moyal formula
\begin{equation}
\label{prodrule1}
C_W\!\left(\bx,\bp
\right)=\,\left.
\exp\!\!\left[ \frac{i\hbar}{2}
\sum_{i=1}^{3}\left(
\frac{\partial}{\partial b_i}\,
\frac{\partial}{\partial z_i} - \frac{\partial}
{\partial a_i}\,\frac{\partial}{\partial y_i}
\right)\right]\! 
A_W(\bz,\ba)
B_W(\by,\bb)
\right|_{
\by=\bz
=\bx
\atop
\ba=\bb
=\bp}\,,
\end{equation}
where the subscript ``$i$'' characterizes the Cartesian
coordinates of a vector, i.e.\ 1,2,3 stands for $x,y,z$
respectively, and the auxiliary vectors 
$\ba,\bb, 
\by,\bz$
specify which of the factors $A_W(\bx,
\bp)$, $B_W(\bx,
\bp)$ has to be differentiated with 
respect to $\bx$
and $\bp$.
Note that the operator in the exponential is just the
ordinary $(\bx,\bp)$
Poisson bracket operator, so that expanding the right
hand side of (\ref{prodrule1}) with respect to $\hbar$
yields
\begin{equation}
\label{prodrule2}
C_W(\bx,\bp)
= A_W(\bx,\bp)
B_W(\bx,\bp)
+ \frac{i \hbar}{2}\{A_W,B_W\} 
+ {\cal O}\!\left(\hbar^2\right).
\end{equation}
Equation (\ref{prodrule1}) may also be interpreted as defining a
bilinear, associative and non-commutative product on the space
of symbols,
\begin{equation}
\label{starproddef1}
C_W(\bx,\bp)
\equiv A_W(\bx,\bp)
* B_W(\bx,\bp)\,,
\end{equation}
denoted as the star product or Weyl product.

Suppose we are given a phase space function of the form
\begin{equation}
\label{fdefps1}
f(\bx,\bp) = x_i^m\,p_j^n\,, 
\end{equation}
with $m,n\in \Natural_0$ (= non-negative integers).
Evaluating equation (\ref{Weylexpansion1}) we obtain the
corresponding operator \cite{McC32}
\begin{equation}
\label{fdefop1}
\hat f\!\left(\hat{\bx},
\hat{\bp}\right) = \frac{1}{2^m}
\sum_{l=0}^{m} {m \choose l}\,\hat x_i^l\,\hat p_j^n\,
\hat x_i^{m-l}\,.
\end{equation}
It can be constructed in the following way: First take $\hat 
x_i$ $m$ times, $\hat p_j$ $n$ times, put them in all
possible permutations with equal weights and divide by the
number of terms.
The result is called the totally symmetrized or Weyl ordered
product, written as $\mbox{Symm}\!\left(\hat x_i^m \hat p_j^n
\right)$.
Finally apply the commutation relation $[\hat x_i,\hat p_j]=
i\hbar\,\delta_{ij}$ to bring the $\hat p_j$'s together at
various positions of the product with no terms proportional
to $\hbar$ remaining.
Due to the linearity of the Weyl transform, equations 
(\ref{fdefps1}) and (\ref{fdefop1}) generalize to any
analytic function on phase space.

So far we have used position $\bx$ 
and canonical momentum $\bp$ as basic variables.
If a magnetic field
$\bB(\bx)
\!=\!\nabla\!\!\times\!\!
\bA(\bx)$
is switched on, the canonical momentum is no longer completely
physical because of its gauge dependence.
However, the operator of the kinetic momentum,
\begin{equation}
\label{hatkdef}
\hat{\bk} \equiv \hat{\bp} - 
\frac{q}{c}\,\,\hat{\!\bA}
(\hat{\bx})
= m\hat{\bv}\,,
\end{equation}
is gauge invariant because its expectation value is
not effected by a gauge transformation \cite{Coh77a}.
In contrast to $\hat{\bp}$ 
the Cartesian components of $\hat{\bk}$
do not commute with one another,
\begin{equation}
\label{commkij}
\left[\hat k_i,\hat k_j\right] = i\,\hbar\,\frac{q}{c}\,
\epsilon_{ijk} B_k(\bx),
\end{equation}
whereas their commutation relations with
$\hat{\bx}$ parallel those for 
$\hat{\bp}$ and $\hat{\bx}$, 
$[\hat x_i,\hat k_j] = i \hbar \delta_{ij}$.

If we replace the canonical momentum $\hat{\bp}$ in
(\ref{Tdef1}) with the kinetic momentum $\hat{\bk}$,
the operators
\begin{equation}
\label{Tdef2}
\hat T\!\left(\bu,\bv
\right)
\,\equiv\,
\exp\!\left[\,i\!\left(
\bu\!\cdot\!\hat{\bk} +
\bv\!\cdot\!\hat{\bx}
\right)
\right]
\end{equation}
become gauge invariant and hence the basic operators
\begin{equation}
\label{Deltadef2}
\hat\Delta\!\left({\bx},
{\bk}\right)
\equiv
\left(\frac{\hbar}{2\pi}\right)^{\!3}\!
\int\!d^3\!\bu\,d^3\!\bv
\exp\!\left[\,i\!\left(
\bu\!\cdot\!\bk +
\bv\!\cdot\!\bx
\right)\right]
\hat T\!\left(-\bu,-\bv
\right)\,,
\end{equation}
are gauge invariant as well.
The Weyl symbol $A_W(\bx,
\bk)$ of an operator $\hat A$ is now defined with respect
to the new basis in the same way as in the field-free
case,
\begin{equation}
\label{Weylsymboldef2}
A_W\!\left(\bx, \bk\right) =
\mbox{Tr}\!\left[ \hat A
\,\hat\Delta\!\left({\bx},
{\bk}\right)\right]\,, 
\end{equation}
or, equivalently,
\begin{equation}
\label{Weylexpansion2}
\hat A = 
\int \frac{d^3\!\bx\,d^3\!\bk}
{h^3}
\,\hat\Delta\!\left({\bx},
{\bk}\right)
A_W\!\left(\bx, \bk\right)\,.
\end{equation}
From equation (\ref{Weylsymboldef2}) and the properties of 
gauge invariant operators (cf.\ \cite{Coh77a}) it is 
obvious that the Weyl symbol of a gauge invariant operator
does not change its value under gauge transformations.
Thus, viewed as a phase space function, the symbol is also
gauge invariant.
According to equation (\ref{Weylexpansion2}) the opposite
is also true:
The operator corresponding to a gauge invariant phase space
function is itself gauge invariant, i.e.\ its mean value does
not change under gauge transformations.

Writing $A_W(\bx,\bk)$ as a Fourier integral,
\begin{equation}
\label{Weylexpansion2alt}
A_W(\bx, \bk) = 
\int\!d^3\bu\,d^3\bv 
\exp\!\left[\,i\!
\left(\bu\!\cdot\!\bk + 
\bv\!\cdot\!\bx
\right)\right] 
\tilde A\!\left(\bu,
\bv\right),
\end{equation}
one can show by inserting definition (\ref{Deltadef2})
into (\ref{Weylsymboldef2}) that the Fourier transform
$\tilde A\!\left(\bu,
\bv\right)$ may also be obtained from 
\begin{equation}
\label{Atildedef2}
\tilde A\!\left(\bu,
\bv\right) = 
\left( \frac{\hbar}{2 \pi}\right)^{\!3}\!
\mbox{Tr}\!\left[ \hat A \,
\hat T\!\left(-\bu,-\bv
\right)\right]\,.
\end{equation}
According to (\ref{Deltadef2}), (\ref{Weylexpansion2}), and 
(\ref{Weylexpansion2alt}) the operator $\hat A$ can similarly
be expressed in terms of $\tilde A\!\left(
\bu,\bv\right)$
via
\begin{equation}
\label{Weylexpansion3}
\hat A =  
\int\!d^3\bu\,d^3\bv 
\tilde A\!\left(\bu,
\bv\right) \hat T(
\bu,\bv)\,.
\end{equation}
Equations (\ref{Tdef2}), (\ref{Weylexpansion2alt})
and (\ref{Weylexpansion3}) are the gauge invariant
generalization of Weyl's original correspondence rule 
\cite{Wey27} for phase space functions and quantum mechanical
operators. 

To derive a product rule for gauge invariant Weyl symbols we
will have to evaluate matrix elements of the form $\langle
\bx'|\hat A |
\bx\rangle$.
For this purpose it is of advantage to express the operator
$\exp(i\,\bu\!\cdot
\!\hat{\bk})$ occurring in $\hat T
(\bu,\bv)$
by the translation operator $\exp(i\,
\bu\!\cdot\!
\hat{\bp})$.
The latter acts on an operator function $\hat f(\hat
{\bx})$ and a position eigenstate
$|\bx\rangle$ in the following way
\begin{eqnarray}
\label{transopac1}
\exp(i\,\bu\!\cdot\!
\hat{\bp})
\hat f(\hat{\bx})&=&
\hat f(\hat{\bx} +
\hbar \bu)
\exp(i\,\bu\!\cdot\!
\hat{\bp})\,, \\
\exp(i\,\bu\!\cdot\!
\hat{\bp})
\,|\bx\rangle
&=&
\label{transopac2}
|\bx-\hbar\bu\rangle.
\end{eqnarray}
As proven in \cite{Ser86}, $\exp(i\,\bu\!
\cdot\!\hat{\bk})$ and 
$\exp(i\,\bu\!\cdot\!
\hat{\bp})$
are related by
\begin{eqnarray}
\label{BCH1}
\exp(i\,\bu\!
\cdot\!\hat{\bk})&=&
\exp(i\,\bu\!
\cdot\!\hat{\bp})
\exp\!\left\{- i\,\frac{q}{c}\,
\bu\!\cdot\!\!\int_0^1\!\!
\,\hat{\!\bA}\!\left(\hat{\bx}
-\hbar\tau\bu
\right)d\tau\right\} \nonumber \\
&=&
\exp\!\left\{- i\,\frac{q}{c}\,
\bu\!\cdot\!\!\int_0^1\!\!
\,\hat{\!\bA}\!\left(\hat{\bx}
+\hbar\tau\bu
\right)d\tau\right\}
\exp\!\left(i\,\bu\!
\cdot\!\hat{\bp}\right).
\end{eqnarray}
Using (\ref{transopac1}), (\ref{BCH1}), and the 
Baker-Campbell-Hausdorff type formula
\begin{eqnarray}
\label{BCH2}
\hat T(\bu,\bv)&=&
\exp(-\,i\,\bu
\!\cdot\!\bv/2)
\exp(i\,\bu\!\cdot\!
\hat{\bk})
\exp\!\left(i\,\bv\!\cdot\!
\hat{\bx}\right) \nonumber \\
&=&
\exp(i\,\bu
\!\cdot\!\bv/2)
\exp\!\left(i\,\bv\!\cdot\!
\hat{\bx}\right)
\exp(i\,\bu\!\cdot\!
\hat{\bk})\,,
\end{eqnarray}
a straightforward calculation shows that the product of
two gauge invariant $\hat T$ operators may be cast into
the form
\begin{eqnarray}
\label{duplication2}
\hat T\!\left(\bu,\bv
\right)
\hat T\!\left(\bu',\bv'
\right)
\,=\,& &\exp\!\left\{i\,\frac{q}{c}\,
\bu\!\cdot\!\!\int_0^1\!
\left[\,\hat{\!\bA}\!\left(\hat{\bx}
+\hbar\tau\!\left(\bu + 
\bu'\right)\right)
- \,\hat{\!\bA}\!\left(\hat{\bx}
+\hbar\tau\bu
\right)\right]d\tau\right\}\!\times 
\nonumber \\
& & \exp\!\left\{\,i\,\frac{q}{c}\,
\bu'\!\cdot\!\!\int_0^1\!\left[
\,\hat{\!\bA}\!\left(\hat{\bx}
+\hbar\tau\!\left(\bu + 
\bu'\right)\right)
- \,\hat{\!\bA}\!\left(\hat{\bx}
+\hbar\bu
+\hbar\tau\bu'\right)\right] d\tau 
\right\}\!\times
\nonumber \\
& &\exp\!\left[\,i\,\frac{\hbar}{2}\,
\left(\bu\!\cdot\!\bv'
-\bu'\!\cdot\!\bv\right)\right]
\hat T\!\left(\bu+\bu',
\bv+\bv'\right),
\end{eqnarray}
which reduces to the ordinary duplication formula
(\ref{duplication1}) if the vector potential $\bA
(\bx)$ vanishes.
Another consequence of equations (\ref{BCH1}) and
(\ref{BCH2}) is that the trace of
$\hat T(\bu,\bv)
\,\hat T(-\bu',-\bv')$
is given by
\begin{equation}
\label{traceTprod}
\mbox{Tr}\left[
\hat T\!\left(\bu,\bv
\right)
\hat T\!\left(-\bu',-\bv'
\right)\right]
=\left(\frac{2\pi}{\hbar}\right)^{\!3}
\delta\!\left(\bu-\bu'
\right)
\delta\!\left(\bv-\bv'
\right)\,.
\end{equation}

After these preliminary remarks we are ready to
determine the Weyl symbol $C_W(\bx,
\bk)$ of the operator product $\hat C = \hat A \hat B$
in terms of $A_W(\bx,\bk)$
and $B_W(\bx,\bk)$.
From equations (\ref{Weylexpansion3}) and 
(\ref{duplication2}) we find that
\begin{eqnarray}
\label{CAtildeBtilde1}
\hat C = \hat A \hat B &=&
\int d^3\bu\,d^3\bv\,
d^3\bu'\,d^3\bv'
\tilde A\!\left(\bu,
\bv\right)
\tilde B\!\left(\bu',
\bv'\right)
\hat T\!\left(\bu,
\bv\right)
\hat T\!\left(\bu',
\bv'\right) \nonumber \\
&=&
\int d^3\bu\,d^3\bv\,
d^3\bU\,d^3\bV 
\tilde A\!\left(\bu,\bv\right)
\tilde B\!\left(\bU-\bu,
\bV-\bv\right)\!\times \nonumber \\ 
& &\hspace{0.2in}
\exp\!\left[\,i\,\frac{\hbar}{2}\,\left(
\bu\!\cdot\!\bV-
\bU\!\cdot\!\bv
\right)\right]
\hat F\!\left(\bu,\bU;
\hat{\bx}\right)
\hat T\!\left(\bU,\bV
\right),
\end{eqnarray}
where we introduced new integration variables
$\bU\!\equiv \bu+
\bu'$, $\bV\!\equiv
\bv +\bv'$
and the function
\begin{eqnarray}
\label{hatFdef}
F\!\left(\bu_1,\bu_2;
\bx\right) \equiv\,& &
\exp\!\left\{i\,\frac{q}{c}\,
\bu_1\!\cdot\!\!\int_0^1\!\!\left[\,
\bA\!\left(\bx
+\hbar\!\left(1-\tau\right)\!\bu_1
+ \hbar\tau\bu_2\right)
- \bA\!\left(\bx
+\hbar\tau\bu_1
\right)\right]d\tau\right. - 
\nonumber \\
& &\hspace{0.35in}\left.i\,\frac{q}{c}\,
\bu_2\!\cdot\!\!\int_0^1\!\!\left[\,
\bA\!\left(\bx+\hbar
\!\left(1-\tau\right)\!\bu_1
+\hbar\tau\bu_2\right)
-\,\bA\!\left(\bx+\hbar\tau
\bu_2\right)\right] d\tau \right\}
\nonumber \\
\equiv\,& &\exp\!\left[\,i\,\frac{q}{c}\,
f\!\left(\bu_1,
\bu_2;\bx\right)\right],
\end{eqnarray}
which is equal to unity in the field-free case
$\bA(\bx)\equiv 0$.
Note that in (\ref{CAtildeBtilde1}) $F$ depends on the
position operator $\hat{\bx}$
and hence is itself an operator.

Inserting (\ref{CAtildeBtilde1}) into
(\ref{Atildedef2}) and making use of (\ref{traceTprod})
and the completeness of the position eigenstates 
$|\bx\rangle$ yields the Fourier transform of
$C_W(\bx,\bk)$,
\begin{eqnarray}
\label{Ctildeprod1}
\tilde C\!\left(\bu,\bv
\right) 
=& &\left(\frac{\hbar}{2\pi}\right)^{\!3}\!\int
d^3\bx\,d^3\bu'
\,d^3\bv'\,d^3\bV 
\tilde A\!\left(\bu',\bv'
\right)
\tilde B\!\left(\bu-\bu',
\bV-\bv'\right)
\exp\!\left[\,i\!\left(\bV-\bv
\right)\!\cdot\!\bx\right]\!\times
\nonumber \\
& &\hspace{0.68in}\exp\!\left\{\,i\,\frac{\hbar}{2}\,
\left[\left(\bu+\bu'
\right)\!\cdot\!\bV-
\bu\!\cdot\!\left(\bv
+\bv'\right)\right]\right\}
F\!\left(\bu',\bu;
\bx\right),
\end{eqnarray}
so that as an intermediate result the Weyl symbol of
$\hat C$ reads
\begin{eqnarray}
\label{CWder1}
C_W\!\left(\bx,\bk
\right) =& &\int\!d^3\bu'\,
d^3\bv'\,d^3\bu''\,d^3\bv'' 
\tilde A\!\left(\bu',\bv'
\right)
\tilde B\!\left(\bu'',\bv''
\right)\exp\!\left\{\,i\left[\left(\bv'
+\bv''\right)\!\cdot\!\bx
+\left(\bu'+\bu''
\right)\!\cdot\!\bk
\right]\right\}\!\times
\nonumber \\
& &\hspace{0.16in}\exp\!\!\left[\,i\,\frac{\hbar}{2}\,
\left(\bu'
\!\cdot\!\bv'' - \bu''
\!\cdot\!\bv'\right)\right]
F\!\!\left(\bu',\bu'+
\bu'';\bx
-\frac{\hbar}{2}\left(\bu'
+ \bu''\right)\right),
\end{eqnarray}
where $\bu''\equiv\bu
-\bu'$, $\bv''
\equiv\bV-\bv'$. 
As a next step we want to express the right hand side of
(\ref{CWder1}) in terms of $A_W\!\left(
\bx,\bk\right)$,
$B_W\!\left(\bx,
\bk\right)$ and their derivatives with respect to
$\bx$ and $\bk$.
A helpful observation is that by setting $\bA\!\left
(\bx\right)\equiv 0$ the above equation reduces to the
one appearing in the derivation of the ordinary product
rule \cite{Bal84}.
There, the factor $\exp\!\left[\,i\hbar
\left(\bu'\!\cdot\!\bv''
- \bu''\!\cdot\!
\bv'\right)\!/2\right]$
is expanded into a power series with respect to $\hbar$.
The variables $u_i',u_i'',v_i',v_i''$ occurring in each
term of this series are generated from
$\exp\!\left\{\,i\left[\left(\bv'
+\bv''\right)\!\cdot\!\bx
+\left(\bu'+\bu''
\right)\!\cdot\!\bk
\right]\right\} =
\exp\left[\,i\!\left(\bv'
\!\cdot\!\bx + \bu'
\!\cdot\!\bk
\right)\right]
\exp\!\left[\,i\!\left(\bv''
\!\cdot\!\bx + \bu''
\!\cdot\!\bk\right)\right]$
by differentiation processes.
This is achieved by replacing the variables $\bx$ and 
$\bk$ with $\bz$ and 
$\ba$ in the first exponential factor and with 
$\by$ and $\bb$
in the second one and then applying the operator 
$i\hbar\left(\partial/\partial\bz
\!\cdot\!\partial/\partial\bb - \partial/
\partial\by\!\cdot\!\partial/\partial
\ba\right)\!/2$ and appropriate powers
of it to the product.
The resulting total differential operator has the form
$\exp\!\left[i\hbar\left(\partial/\partial
\bz\!\cdot\!\partial/\partial
\bb - \partial/\partial
\by\!\cdot\!\partial/\partial
\ba\right)\!/2\right]$.
If it is taken outside of the integral, the latter may
be evaluated and one finally gets Moyal's formula
(\ref{prodrule1}).

To employ a comparable algorithm in the case of
non-zero vector potential we have to extract 
$\bu'$ and  $\bu''$ 
from the integrals in 
$F\!\left(\bu',\bu'
+\bu'';\bx
-\hbar(\bu'+\bu'')\!/2
\right)$. 
For this purpose the vector potentials occurring in the
exponent of (\ref{hatFdef}) are expanded into Taylor series
around the position $\bx$.
After some additional algebraic manipulations we obtain
\begin{eqnarray}
\label{fexp1}
f\!\left(\bu',\bu'\!\!\!
\right.&+&\left.\!
\bu'';\bx
-\hbar\left(\bu'
+ \bu''\right)\!/2\right) =
\sum_{n=0}^{\infty}\,\frac{\hbar^n}{n!} \sum_{i_1,\dots,i_n,l=1}^{3}
\frac{\partial^n\! A_l}{\partial x_{i_1}
\partial x_{i_2}\dots\partial x_{i_n}}\times
\nonumber \\
& &\hspace{-0.29in}\left\{
-\,u_l'\int_0^1\! \prod_{j=1}^{n}\!
\left[\,\left(1/2 - \tau\right)u_{i_j}' - 1/2\,u_{i_j}''\right]\!
d\tau\,+\,u_l''\int_0^1\!\prod_{j=1}^{n}\!\left[\,1/2\,u_{i_j}'
+ \left(1/2 - \tau\right) u_{i_j}''\right]\!d\tau\right.
\nonumber \\
&-&\left.\left(u_l'+u_l''\right)
\prod_{j=1}^{n}(u_{i_j}'+u_{i_j}'')
\int_0^1\!\!\left(\tau-1/2\right)^n\!d\tau
\right\}.
\end{eqnarray}
Noting that for $k\in \Natural_0$
\begin{equation}
\label{tauint}
\int_0^1\! (\tau-1/2)^kd\tau = \left(-\frac{1}{2}\right)^{k+1}
\frac{(-1)^{k+1}-1}{k+1} = 
\left\{ \begin{array}{ll} 0 & \mbox{, if $k=2m+1,\;m\in\Natural_0$,}
\\ \left(-\frac{1}{2}\right)^k\frac{1}{k+1} > 0 & \mbox{, if $k=2m,
\;m\in\Natural_0$,} \end{array} \right.
\end{equation}
we arrive at
\begin{eqnarray}
\label{fexp2}
& &f\!\left(\bu',\bu+
\bu'';\bx
-\hbar\left(\bu'
+ \bu''\right)\!/2\right)=\,
\sum_{n=1}^{\infty}\,\frac{\hbar^n}{n!}
\left(-\frac{1}{2}\right)^{n+1}\!\frac{1}{(n+1)^2}\!\!
\sum_{r,j,l,\atop i_1,\dots,i_{n-1}=1}^{3}\!\!
\!\epsilon_{jlr}\,\frac{\partial^{n-1}B_r}
{\partial x_{i_1}\dots\partial x_{i_{n-1}}}\times
\nonumber \\
& &\hspace{0.25in}u_j'\,u_l''\,
\sum_{k=1}^{n} {n+1 \choose k} \left[\left(1-(-1)^k\right)
(n+1) - \left(1 - (-1)^{n+1}\right) k\,\right]
u_{i_1}'\dots u_{i_{k-1}}'u_{i_k}''\dots u_{i_{n-1}}''\,.
\end{eqnarray}

The most important result of the preceding calculation is
that the expansion of $f$ includes only derivatives of the
magnetic field $\bB(\bx)$.
Physically this was to be expected because the product of
two gauge invariant symbols is itself gauge invariant.
Therefore the integrand on the right hand side of
(\ref{CWder1}) must not depend on the chosen gauge.
As all other factors satisfy this condition,
the function $F\!\left(\bu',
\bu'+\bu'';
\bx-\hbar\left(\bu'
+ \bu''\right)\!/2\right)$ has to be 
gauge invariant as well.
This is certainly true if it is a functional of the magnetic
field. 

Now we continue just like in the field-free case.
Writing
\begin{equation}
\label{expprod1}
\exp\!\left\{\,i\!\left[\left(\bv'
+\bv''\right)\!\cdot\!\bx
+\left(\bu'+\bu''
\right)\!\cdot\!\bk
\right]\right\} =
\left.
\exp\left[\,i\!\left(\bv'
\!\cdot\!\bz + \bu'
\!\cdot\!\ba
\right)\right]
\exp\!\left[\,i\!\left(\bv''
\!\cdot\!\by + \bu''
\!\cdot\!\bb\right)\right]
\right|_{
\by=\bz=
\bx
\atop
\ba=\bb=
\bk}
\end{equation}
we may generate the variables $u_i',u_i'',v_i',v_i''$
occurring in an analytic function which is multiplied to the
right by this exponential by differentiating the latter with
respect to the auxiliary variables $a_i,b_i,y_i,z_i$.
This is formally equivalent to substituting 
\begin{equation}
\label{substitution1}
u_i'\rightarrow -\,i\,\frac{\partial}{\partial a_i}\,,\;
u_i''\rightarrow -\,i\,\frac{\partial}{\partial b_i}\,,\;
v_i'\rightarrow -\,i\,\frac{\partial}{\partial z_i}\,,\;
v_i''\rightarrow -\,i\,\frac{\partial}{\partial y_i}
\end{equation}
in the analytic function.
Hence, if we introduce gauge invariant operators
\begin{eqnarray}
\label{Lopdef}
{\cal L}&\equiv&\frac{1}{2}\,\sum_{i=1}^{3}
\frac{\partial}{\partial a_i}\,
\frac{\partial}{\partial y_i} -
\frac{\partial}{\partial b_i}\,
\frac{\partial}{\partial z_i}, \\
\label{Lnopdef}
{\cal L}_n\,&\equiv&\,\left(\frac{i}{2}\right)^{n+1}\!
\frac{1}{(n+1)^2\, n!}\!\!
\sum_{r,j,l,\atop i_1,\dots,i_{n-1}=1}^{3}\!
\!\epsilon_{jlr}\,\frac{\partial^{n-1}B_r}
{\partial x_{i_1}\dots\partial x_{i_{n-1}}}\,
\frac{\partial}{\partial a_j}\,
\frac{\partial}{\partial b_l}\,
\sum_{k=1}^{n} {n+1 \choose k}\times
\nonumber \\
& &\hspace{0.06in} \left[\left(1-(-1)^k\right)
(n+1) - \left(1 - (-1)^{n+1}\right) k\,\right] 
\frac{\partial}{\partial a_{i_1}}
\dots\frac{\partial}{\partial a_{i_{k-1}}}\,
\frac{\partial}{\partial b_{i_k}}
\dots \frac{\partial}{\partial b_{i_{n-1}}}\,,
\end{eqnarray}
$n \in \Natural$, and leave aside questions of convergence,
we may write
\begin{eqnarray}
\label{funcoprel1}
& &\hspace{3ex}
\exp\!\left[\,i\,\hbar
\left(\bu'
\!\cdot\!\bv'' - \bu''
\!\cdot\!\bv'\right)/2\right]
\exp\!\left\{\,i\left[\left(\bv'
+\bv''\right)\!\cdot\!\bx
+\left(\bu'+\bu''
\right)\!\cdot\!\bk
\right]\right\}=\nonumber \\
& &\hspace{5ex}
\exp\!\left[ 
- i \hbar\,{\cal L}\right]
\left.
\exp\!\left[\,i\!\left(\bv'
\!\cdot\!\bz + \bu'
\!\cdot\!\ba
\right)\right]
\exp\!\left[\,i\!\left(\bv''
\!\cdot\!\by + \bu''
\!\cdot\!\bb\right)\right]
\right|_{
\by=\bz=
\bx
\atop
\ba=\bb=
\bk},
\end{eqnarray}
and
\begin{eqnarray}
\label{funcoprel2}
& &F\!\left(\bu',\bu'+
\bu'';\bx
-\hbar\left(\bu'
+ \bu''\right)\!/2\right)
\exp\!\left\{\,i\left[\left(\bv'
+\bv''\right)\!\cdot\!\bx
+\left(\bu'+\bu''
\right)\!\cdot\!\bk
\right]\right\}=
\nonumber \\
& &\hspace{0.5ex}
\exp\!\left[
- i\,\frac{q}{c}
\sum_{n=1}^{\infty} \hbar^n {\cal L}_n \right]
\left.
\exp\!\left[\,i\!\left(\bv'
\!\cdot\!\bz + \bu'
\!\cdot\!\ba
\right)\right]\!
\exp\!\left[\,i\!\left(\bv''
\!\cdot\!\by + \bu''
\!\cdot\!\bb\right)\right]
\right|_{
\by=\bz=
\bx
\atop
\ba=\bb=
\bk}.
\end{eqnarray}

All operators ${\cal L}$ and ${\cal L}_n$, $n\!\in\!\Natural$, 
commute with one another because they contain $a_i,b_i,y_i,
z_i$ only as differentiating variables (the magnetic field
$\bB(\bx)$
and its derivatives occurring in ${\cal L}_n$ depend on the
position $\bx$ and are hence not effected 
by a differentiation with respect to these variables).
Therefore, the integrand in (\ref{CWder1}) can be generated
by the action of 
\begin{equation}
\label{Pdeforig}
{\cal P}\!\equiv\!\exp\!\left[ 
- i\, \hbar\,{\cal L} - i\,(q/c)
\sum_{n=1}^{\infty} \hbar^n {\cal L}_n \right]
\end{equation}
on $\left.\exp\left[\,i\!\left(\bv'
\!\cdot\!\bz + \bu'
\!\cdot\!\ba
\right)\right]
\exp\!\left[\,i\!\left(\bv''
\!\cdot\!\by + \bu''
\!\cdot\!\bb\right)\right]
\right|_{
\by=\bz=
\bx
\atop
\ba=\bb=
\bk}$.
Taking the total differential operator outside of the
integral (\ref{CWder1}) we finally obtain
\begin{equation}
\label{Moyalrule1}
C_W\!\left(\bx,\bk\right)=\,\left.
\exp\!\left[ - i\, \hbar\,{\cal L} - i\,\frac{q}{c}
\sum_{n=1}^{\infty} \hbar^n {\cal L}_n \right]\!
A_W(\bz,\ba)
B_W(\by,\bb)\right|_{
\by=\bz=
\bx
\atop
\ba=\bb=
\bk}
\,\equiv\,\left[A_W*B_W\right]\!\left(\bx,\bk\right) ,
\end{equation}
which is the generalization of Moyal's formula to gauge
invariant Weyl symbols.
As in the case of the ordinary Weyl transform one can
show that the star product defined by (\ref{Moyalrule1})
is bilinear and associative.

Before turning to the semiclassical analysis of guiding
center motion, let us investigate equation
(\ref{Moyalrule1}) in more detail.
Expanding the exponential operator $\cal P$ into a power
series with respect to $\hbar$ yields
\begin{eqnarray}
\label{Moyalprodexp}
& &{\cal P} =
1 - i\,\hbar \left( {\cal L} + \frac{q}{c}\,{\cal L}_1
\right) - \hbar^2 \left[ \frac{1}{2}\,\left( {\cal L} +
\frac{q}{c}\,{\cal L}_1 \right)^2 + i\,\frac{q}{c}\,
{\cal L}_2\right]
\nonumber \\
& &\hspace{4.5ex}
+\,i\,\hbar^3\left[ \frac{1}{6} \left({\cal L} +
\frac{q}{c}\,{\cal L}_1\right)^3 + i\,\frac{q}{c}\,
\left( {\cal L} + \frac{q}{c}\,{\cal L}_1
\right)\!{\cal L}_2 - \frac{q}{c}\,{\cal L}_3 \right] +
{\cal O}(\hbar^4)\,.
\end{eqnarray} 
The second term on the right hand side of (\ref{Moyalprodexp})
is equal to  $i\hbar/2$ times the
$(\bx,\bp)$
Poisson bracket operator.
This is seen most easily by expressing the Poisson
bracket of two arbitrary phase space functions in terms
of $\bx$- and $\bk$-derivatives,
\begin{equation}
\label{Poissonchain}
\left\{f,g\right\} = \sum_{i=1}^{3} \frac{\partial f}
{\partial x_i}\frac{\partial g}{\partial k_i} 
- \frac{\partial f}{\partial k_i}\frac{\partial g}
{\partial x_i}\, + \frac{q}{c}
\sum_{j,l,r=1}^{3} \epsilon_{jlr} 
\frac{\partial f}{\partial k_j}\frac{\partial g}
{\partial k_l}\,B_r\,.
\end{equation}
A comparison of (\ref{prodrule2}) and
(\ref{Moyalprodexp}) shows that the first order terms of
both expansions coincide.
However, the higher order terms in (\ref{prodrule2})
turn out to be gauge dependent and hence differ from those
in (\ref{Moyalprodexp}).

Using their definitions (\ref{Lopdef}), (\ref{Lnopdef}), one 
can derive the following symmetry properties of the operators
${\cal L}$, ${\cal L}_n$, 
\begin{eqnarray}
\label{Lsymm}
{\cal L}^m\!\left(B_W, A_W\right)&=&\,(-1)^m{\cal L}^m
\!\left(A_W, B_W\right),
\\
\label{Lnsymm}
{\cal L}_n^m \!\left( B_W, A_W \right)&=&\,
(-1)^{nm}{\cal L}_n^m\!\left(A_W, B_W\right),\;m\!\in\!\Natural\,, 
\end{eqnarray}
which cause the star product to be
non-commutative.
The difference
\begin{equation}
\label{Moybrack}
A_W*B_W - B_W*A_W\,\equiv\, \left[ A_W, B_W\right]_M
\end{equation}
is called the Moyal bracket of $A_W$ and $B_W$.
According to (\ref{Moyalprodexp}) it may be expanded into
\begin{eqnarray}
\label{Moyalbrackexp}
\left[ A_W, B_W\right]_M 
\equiv {\cal M}\left(A_W B_W\right) &=& \left\{
- 2 i\,\hbar \left( {\cal L} + \frac{q}{c} {\cal L}_1 
\right) +\,2\,i\,\hbar^3\left[\,\frac{1}{6} \left({\cal L}  
+ \, \frac{q}{c} {\cal L}_1\right)^3 
\right. \right. \nonumber \\ 
& &\hspace{2ex} \left.\left. 
+\, i\,\frac{q}{c}
\left( {\cal L} + \frac{q}{c} {\cal L}_1 
\right)\!
{\cal L}_2 - \frac{q}{c}{\cal L}_3 \frac{}{}\right] 
+ {\cal O}(\hbar^5)\right\}A_W B_W\,,
\end{eqnarray}
where the leading order term is just $i\,\hbar
\left\{A_W,B_W\right\}$.
In contrast to the ordinary Weyl calculus the
gauge invariant Moyal bracket operator $\cal M$ cannot be
written in closed form.

Finally one can prove from (\ref{Moyalrule1}) by induction
that the Weyl symbol of the operator
\begin{equation}
\label{Weylsymbopgiex1}
\hat A = 
\mbox{Symm}\!\left(\hat k_{i_1}\hat k_{i_2} \dots 
\hat k_{i_n} \right),\;
i_j\in\{1,2,3\},\;1\leq j\leq n\,,
\end{equation}
is given by
\begin{equation}
\label{Weylsymbgiex1}
A_W = k_1^{n_1}\,k_2^{n_2}\,k_3^{n_3},\;
n_1+n_2+n_3 = n \,,
\end{equation}
if $\hat k_i,\,1\!\leq i\!\leq 3$, appears $n_i$ 
times in the operator product $\hat k_{i_1}\hat 
k_{i_2} \dots \hat k_{i_n}$.
The relation above is a direct consequence of the
non-commutativity of the Cartesian components of 
$\hat{\bk}$.
In general, if $f(\bx)$ is an analytic function of
$\bx$, the operator related to
\begin{equation}
\label{Weylsymbgiex2}
A_W = f(\bx)\,k_1^{n_1}\,k_2^{n_2}\,k_3^{n_3}
\end{equation}
reads
\begin{equation}
\label{Weylsymbopgiex2}
\hat A \equiv \mbox{Symm}_{\{\hat k_{i_j}\}}\!
\left(\,\frac{1}{2} \sum_{l=0}^{n} {n \choose l}\,
\hat k_{i_1}\hat k_{i_2} \dots \hat k_{i_{n-l}} \, 
\hat f(\hat{\bx}) \, \hat k_{i_{n-l+1}} 
\hat k_{i_{n-l+1}} \dots \hat k_{i_n} \right),
\end{equation}
where $n\!=\!n_1+n_2+n_3$ and
$\mbox{Symm}_{\{\hat k_{i_j}\}}$ denotes symmetrization
with respect to the operators $\hat k_{i_j}$.

\section*{3.\ Semiclassical description of guiding center motion}

\label{secIII}

We will now apply the gauge invariant Weyl formalism to
describe the motion of a charged particle of mass $m$ in
a strong time-independent magnetic field 
$\bB(\bx)\!=\!
\nabla\!\times\!
\bA(\bx)$
and an additional electrostatic field
$\bE(\bx)\!=\!
-\nabla\phi(\bx)$
semiclassically.
The influence of the latter has not been taken into account
in quantum mechanical calculations \cite{Mar96b,Mar97} so
far. 
Assuming that the guiding center approximation is valid,
we are in particular interested in the lowest order quantum
mechanical correction to the guiding center Hamiltonian.

To incorporate the guiding center approximation into
our theory, we follow the classical calculations and introduce 
an adiabatic parameter $\epsilon$ by replacing the electric
charge $q$ of the particle by $q/\epsilon$ \cite{Lit79,Lit81a},
\begin{equation}
\label{adscal}
q\rightarrow\frac{q}{\epsilon}\;.
\end{equation}
Physical results are recovered at the end of our calculation
by setting $\epsilon\!=\!1$.
In guiding center approximation we are speaking of the
order of an expression in terms of its behavior as $\epsilon\!
\rightarrow\!0$.
The physical meaning and mathematical details of this
limit are discussed in greater detail in \cite{Nor63,Kru65}.
We adopt the convention that the particle variables 
$\bx$ and $\bv$
as well as the fields
$\bA$ and $\bB$
are held constant in this limiting process, i.e.\ are
independent of $\epsilon$.
Since the guiding center approximation breaks down when the
component $E_\parallel$ 
of the electric field parallel to
$\bB$ is of the same magnitude as 
$|\bB|$,
we take $E_\parallel={\cal O}(\epsilon)$ \cite{Lit81a}.

For reasons of notational convenience we will further on
suppress the constants $q,m,c$, which is equivalent to the
following scaling of physical quantities,
\begin{eqnarray}
\label{physcal}
& &\bx \rightarrow 
\frac{1}{\sqrt{m}}\,\bx,\;
\bp \rightarrow 
\sqrt{m}\,\bp,\;
\bv \rightarrow 
\frac{1}{\sqrt{m}} \,\bv,
\nonumber \\
& &\Phi \rightarrow 
\frac{1}{q} \,\Phi,\;
\bA \rightarrow 
\frac{\sqrt{m}c}{q}\,\bA,\;
\bB \rightarrow 
\frac{m c}{q} \,\bB,\;
\bE \rightarrow 
\frac{\sqrt{m}}{q}\, \bE\,.
\end{eqnarray}
The corresponding backward transformations restore the
correct physical units in our results.
Note that with respect to this scaling particle velocity
and kinetic momentum are equal.

Due to the foregoing conventions, the operator $\cal P$
introduced in (\ref{Moyalrule1}) takes the form
\begin {equation}
\label {Ptildedef}
\tilde{\cal P}\equiv\exp\!\left[-i\,\hbar\,
\tilde{\cal L} - \frac{i}{\epsilon}
\sum_{n=1}^{\infty} \hbar^n \tilde{\cal L}_n\right],
\end{equation}
where $\tilde{\cal L},\tilde{\cal L}_n$ denote the
scaled versions of ${\cal L},{\cal L}_n$.
Taking into account the definitions (\ref{Lopdef}),
(\ref{Lnopdef}) of ${\cal L},{\cal L}_n$ the first three
terms of the power series expansion of $\tilde{\cal P}$ with 
respect to $\hbar$ read
\begin{eqnarray}
\label{prodexpscal1}
& &\tilde{\cal P} =
1 - \frac{i\,\hbar}{2}\left[
\sum_{i=1}^{3}\left(
\frac{\partial^2}{\partial a_i\partial y_i} -
\frac{\partial^2}{\partial b_i\partial z_i}\right)
- \frac{1}{\epsilon} \sum_{j,l,r=1}^{3}\epsilon_{jlr}
B_r \frac{\partial^2}{\partial a_j \partial b_l}\right]
\nonumber \\
& &\hspace{3ex}-\,\frac{\hbar^2}{4}\left\{\,\frac{1}{2}
\,\sum_{i,j=1}^{3}
\left( \frac{\partial^4}{\partial a_i \partial a_j 
\partial y_i \partial y_j}\,-\,2 \, \frac{\partial^4}
{\partial a_i \partial b_j \partial y_i \partial z_j}
\,+\,\frac{\partial^4}
{\partial b_i \partial b_j \partial z_i \partial z_j}
\right)\right.\nonumber \\
& &\hspace{4.5ex}\left.
-\,\frac{1}{\epsilon}\sum_{i,j,l,r=1}^{3}\!\!\!\epsilon_{jlr}
\!\left[B_r\left(\frac{\partial^4}{\partial a_i \partial
a_j\partial b_l \partial y_i}\,-\, \frac{\partial^4}
{\partial a_j \partial b_i \partial b_l\partial z_i}
\right) +\,\frac{1}{3}\,\frac{\partial B_r}
{\partial x_i}\!\left(\frac{\partial^3}{\partial a_j 
\partial a_i \partial b_l} - \frac{\partial^3}
{\partial a_j \partial b_l \partial b_i}\right)\right]
\right.\nonumber \\
& &\hspace{4.5ex}\left.
+\,\frac{1}{2}\,\frac{1}{\epsilon^2}
\sum_{j,l,r\atop k,m,s=1}^{3}\epsilon_{jlr}\epsilon_{kms}
\right.\left.B_r\,B_s \,\frac{\partial^4}
{\partial a_j \partial a_k \partial b_l \partial b_m}
\,\right\} + \,{\cal O}\!\left(\hbar^3\right).
\end{eqnarray} 
Since each of the operators $\tilde{\cal L}_n$ has
$\epsilon^{-1}$ attached to it, a simple reasoning shows
that the term proportional to $\hbar^n$ in the expansion
of $\tilde{\cal P}$ includes terms of all orders in
$\epsilon^{-1}$ from $0$ to $n$.
Therefore we may formally write
\begin{equation}
\label{Pcalformdef}
\tilde{\cal P} \equiv\,
\sum_{n=0}^{\infty}\hbar^n{\cal P}_n\,\equiv\,
\sum_{n=0}^{\infty}\,\sum_{m=-n}^{0}\hbar^n\epsilon^{m}
{\cal P}_{n,m}\,. 
\end{equation}
The operators ${\cal P}_{n,m}$ contain derivatives of
order $n\!-\!|m|$ with respect to the position variables
$\by,\bz$
and derivatives of order $n\!+\!|m|$
with respect to the kinetic momentum (=velocity) variables
$\ba,\bb$.
As a consequence of the symmetry properties of
$\tilde{\cal L},\tilde{\cal L}_n$ no terms of even power
in $\hbar$ occur in the expansion of the (scaled) Moyal
bracket operator $\tilde{\cal M}$,
\begin{equation}
\label{Moyalbrexp1}
\tilde {\cal M}= 2 \sum_{n=0}^{\infty}\hbar^{2n+1}
{\cal P}_{2n+1} = 2 \sum_{n=0}^{\infty}\,\sum_{m=-(2n+1)}^{0}
\hbar^{2n+1}\epsilon^{m}
{\cal P}_{2n+1,m}\,.
\end{equation}

Suppose now we are given symbols of the form
\begin{equation}
\label{weylsymbformdef}
A_W\,=\,\sum_{n=0}^{\infty} \hbar^n A_n 
\,=\sum_{m,n=0}^{\infty} \! \hbar^n \epsilon^m A_{n,m}\,,
\;\;\;B_W\,=\,\sum_{n=0}^{\infty} \hbar^n B_n 
\,= \sum_{m,n=0}^{\infty}\! \hbar^n \epsilon^m B_{n,m}\,,
\end{equation}
with the coefficients being analytical functions of
$\bx$ and $\bk$.
From (\ref{Pcalformdef}) and (\ref{weylsymbformdef}) we
obtain the following expansion for the star product of $A_W$
and $B_W$,
\begin{equation}
\label{weylsybprodheps}
A_W*B_W\,=\,\sum_{n=0}^{\infty}\,\sum_{i=0}^{n}\,
\sum_{m=-i}^{\infty}\,
\sum_{l=m}^{m+i} \hbar^n \epsilon^m {\cal P}_{i,m-l}
\sum_{r=0}^{n-i}\,\sum_{s=0}^{l} A_{r,s}B_{n-i-r,l-s}\,.
\end{equation}
A similar calculation yields for the Moyal bracket of
$A_W$ and $B_W$
\begin{eqnarray}
\label{Moyalbrexp2}
[A_W,B_W]_M&=&\sum_{n=0}^{\infty}\hbar^n\left[
\sum_{m=0}^{n} \left[1-(-1)^m\right]{\cal P}_m \left(\,
\sum_{l=0}^{n-m}A_l\,B_{n-m-l}\right)\right]\nonumber \\
&=&\sum_{n=0}^{\infty}\,\sum_{i=0}^{n}\,
\sum_{m=-i}^{\infty}\,
\sum_{l=m}^{m+i} \hbar^n \epsilon^m 
\left[1-(-1)^i\,\right]{\cal P}_{i,m-l}
\sum_{r=0}^{n-i}\,\sum_{s=0}^{l} A_{r,s}B_{n-i-r,l-s}\,.
\end{eqnarray}

Let us now determine the symbols of both guiding center
coordinates and the guiding center Hamiltonian.
The basic features of our method become most transparent
if the direction of the magnetic field is constant with the
electric field being perpendicular to it.
In addition, this case is notationally easier to handle than
the more general one in which the directions of both fields
are varying arbitrarily.
Therefore, we will consider in the following a charged
particle in a magnetic field
$\bB=B(x,y)\be_z$
and electrostatic potential  $\phi(x,y)$ neglecting its motion
parallel to $\bB$.
This kind of planar motion in a strong two-dimensional
magnetic field (i.e.\ of constant direction) is intensively
studied in the context of the Quantum Hall effect \cite{Pra90}.
There, the electric field is weak compared to
$\bB$, which in our
scaling is equivalent to assuming that
$\bE$ is of order $\epsilon$,
$\bE={\cal O}(\epsilon)$.
Classically, this means that the 
$\bE\!\times\!\bB$
drift is of the same order of magnitude as the
$\nabla B$ drift \cite{Lit83}.
Using scaled velocity operators $\hat v_i\!=\!\hat p_i -
\epsilon^{-1}A_i\!=\!\hat k_i, i=1,2$, 
the Hamiltonian in such a field configuration reads
\begin{equation}
\label{hamop}
\hat H =\frac{1}{2}\left(\,\hat v_x^2+ \hat v_y^2\,\right)
+ \hat \phi(\hat x,\hat y)\,.
\end{equation}
Its Weyl symbol is obtained by replacing  operators with their
corresponding phase space functions, taking into account that
the symbol of $\hat v_i^2\!=\!\hat v_i \hat v_i$ is equal to
$v_i\!*\!v_i$,
\begin{equation}
\label{hamopsymb}
H_W =\frac{1}{2}\left(\,v_x*v_x + v_y*v_y\right)
+ \phi(x,y)\,.
\end{equation}

In the special case of a homogeneous magnetic field
$\bB\!=\!B\be_z$,
it is well known that the operators
\begin{equation}
\label{homhatVxVy}
\hat V_x=B^{-1/2}\hat v_x\;,\;\;\hat V_y=B^{-1/2}\hat v_y
\end{equation}
are canonically conjugate and the Hamiltonian has the form
of a one-dimensional harmonic oscillator.
Physically, $\hat V_x$ and $\hat V_y$ describe the gyration
around the magnetic field lines.
To get a complete set of conjugate operators including 
$\hat V_x$ and $\hat V_y$, one has to replace the particle
coordinates with the operators
\begin{equation}
\label{homBXY}
\hat X\!=\!\hat x + 
\epsilon B^{-1} \hat v_y\;,\;\;\hat Y\!=\!\hat y -
\epsilon B^{-1}\hat v_x
\end{equation}
of the guiding center position.
In equations (\ref{homhatVxVy}) and (\ref{homBXY}) questions of
ordering need not be taken into consideration because $B$ is
a real valued constant.
The non-vanishing commutators of $\hat X,\hat Y, \hat V_x,
\hat V_y$ are
\begin{equation}
\label{hommagf}
[\hat V_x,\hat V_y] = i\,\frac{\hbar}{\epsilon}\;,\;\;
[\hat X, \hat Y] = i\,\frac{\hbar\epsilon}{B}\,.
\end{equation}
The Weyl symbols of these operators are obtained by replacing
$\hat x,\hat y,\hat v_x,\hat v_y$ in (\ref{homhatVxVy}) and
(\ref{homBXY}) with the corresponding phase space functions.
We denote them as guiding center symbols and -- leaving away
the subscript ``W'' -- write $X,Y,V_x,V_y$ or 
$\bX,\bV$ for them. 
Their Moyal brackets resemble the commutators (\ref{hommagf})
of the related operators.

Generalizing the results for the homogeneous field to the case
of an arbitrary two-dimensional magnetic field we are looking
for a set of symbols ($X,Y,V_x,V_y$) whose non-vanishing Moyal
brackets are given by
\begin{equation}
\label{Moybrackrel1}
[V_x,V_y]_M= i\,\frac{\hbar}{\epsilon}\;,\;\;
[X,Y]_M= i\,\frac{\hbar\epsilon}{B(X,Y)}\,.
\end{equation}
Their different orders with respect to $\epsilon$ indicate
the different time scales of motion.
They are separated because the symbols $X,Y$ of
the guiding center position commute with those of the
gyration velocity $V_x,V_y$.
The latter are again conjugate to one another.

Concerning the Moyal bracket of the guiding center position
components $X$ and $Y$ two remarks are necessary:
First, the symbol $B(X,Y)$ specifies the strength of the
magnetic field at the position of the guiding center.
The corresponding operator is uniquely determined if $X$ and
$Y$ are expressed in terms of the particle coordinates 
$x,y,v_x,v_y$ and the correspondence rule (\ref{Weylsymbgiex2}),
(\ref{Weylsymbopgiex2}) for arbitrary Weyl symbols and their
operators is applied.
Second, one could think of replacing $X,Y$ with Euler
potentials $\mbox{x}^1(X,Y), \mbox{x}^2(X,Y)$ \cite{Ste70},
thus obtaining a set of conjugate variables to describe guiding
center motion.
However, Euler potentials are non-physical in the same sense
as the vector potential $\bA$ is.
Moreover, in a three-dimensional magnetic field
we get four non-canonical guiding center coordinates instead
of $X$ and $Y$ \cite{Lit81a}.
To transform them into two pairs of canonically conjugate
variables one has to find functions which are less familiar
than Euler potentials and much more difficult to construct.
Therefore we will keep using non-canonical coordinates $X,Y$
to specify the position of the guiding center.

From classical guiding center theory \cite{Lit79} it is
well known that ($X,Y,V_x,V_y$) can be chosen in such a way
that $J\!\equiv\!V_x^2\!+\!V_y^2$ is a constant
of motion which may be interpreted as the generalized
magnetic moment of gyration.
As a direct consequence of the relations (\ref{Moybrackrel1})
the guiding center Hamiltonian must therefore depend on
$V_x,V_y$ only by means of $J$ and its powers.
Accordingly, to find an appropriate set of guiding center
symbols we have to proceed as follows:
First we determine symbols $X,Y,V_x,V_y$ satisfying
(\ref{Moybrackrel1}).
Next we express the particle phase space coordinates
($x,y,v_x,v_y$) in terms of them and insert our result
into the Hamiltonian (\ref{hamopsymb}).
If the latter contains $V_x,V_y$ in other combinations than
$J$ we have to transform to a new set of averaged guiding
center symbols ($\bar{\bX},\bar{\bV}$) satisfying the same
Moyal bracket relations, but $H_W$ depending on the gyration
velocities only via $\bar V_x*\bar V_x\!+\!\bar V_y*\bar V_y$.
This symbol transformation is an analog of the near-identity
Lie transform carried out in the classical calculation
\cite{Lit79,Lit81a}. 

To begin with, let us analyze the Moyal bracket relations
(\ref{Moybrackrel1}) in more detail.
Assuming that the guiding center symbols can be expanded
into power series with respect to $\hbar$ and $\epsilon$
as specified in (\ref{weylsymbformdef}), the Moyal brackets
take the form (\ref{Moyalbrexp2}) with almost all coefficients
vanishing.
Only those of $\hbar\epsilon^{-1}$ and $\hbar\epsilon$ are
different from zero if $A\!=\!V_x, B\!=\!V_y$ and $A\!=\!X,
B\!=\!Y$, respectively.
As stated earlier, the $\hbar$-term of the Moyal bracket 
$[A_W,B_W]_M$ is proportional to the Poisson bracket of $A_0$
and $B_0$.
Therefore, the zero order terms (with respect to $\hbar$) of
the guiding center symbols satisfy the Poisson bracket
relations of classical guiding center theory.
Hence we identify them with the classical guiding center
coordinates, in agreement with the fact that in the limit
$\hbar\!\rightarrow\!0$ Weyl symbols become classical
functions \cite{Lea68a}.
As we will refer to them frequently in the remainder of this
section, the results of the classical guiding center theory
in a two-dimensional magnetic field (using Cartesian
coordinates) are briefly summarized in the appendix.

Concerning higher order terms of the $\hbar$-expansion of the
guiding center symbols one can show that
\begin{equation}
\label{A2n1B2n1zero}
A_{2n+1}=0\;,\;\;B_{2n+1}=0\;,\;\;n\in\Natural_0\;,
\end{equation}
which means that only even powers in $\hbar$ occur.
In addition, one can prove that the coefficients $A_{2n,i}$
are zero for $0\!\leq\!i\!\leq\!2n-1$.
For $i\!\geq\!2n$ they turn out to be homogeneous polynomials
of degree $i\!-\!2n\!+\!1$ for the components of $\bV$ and of
degree $i\!-\!2n$ for the components of $\bX$.
Hence, the expansions of the guiding center symbols with
respect to $\hbar$ and $\epsilon$ take the form
\begin{eqnarray}
\label{Vgciform}
V_i\,&=&\sum_{m,n=0}^{\infty}\!\hbar^{2n}\epsilon^{2n+m}\!
\sum_{k_1+k_2\atop=m+1}\!
\,v_x^{k_1}\,v_y^{k_2}\,{V_i}_{\,2n,2n+m}^{(k_1,k_2)}(x,y)\,,
\\
\label{Xgciform}
X_i&=&\sum_{m,n=0}^{\infty}\hbar^{2n}\epsilon^{2n+m}
\sum_{k_1+k_2\atop=m}
\,v_x^{k_1}\,v_y^{k_2}\,{X_i}_{\,2n,2n+m}^{(k_1,k_2)}(x,y)\,,
\end{eqnarray}
where $i\!=\!1,2$ denotes the Cartesian components of
$\bV$ and $\bX$.
The coefficient functions ${V_i}_{\,2n,2n+m}^{(k_1,k_2)}(x,y),
{X_i}_{\,2n,2n+m}^{(k_1,k_2)}(x,y)$ in (\ref{Vgciform}),
(\ref{Xgciform}) are functionals of the electric and magnetic
fields and thus gauge invariant.

To express the Hamiltonian (\ref{hamopsymb}) in terms of the
guiding center symbols we have to find the corresponding
backward transformations.
Formally they are given by
\begin{eqnarray}
\label{viform}
v_i\,&=&\sum_{m,n=0}^{\infty}\hbar^{2n}\epsilon^{2n+m}
\sum_{k_1+k_2\atop=m+1}
\,V_x^{k_1}\,V_y^{k_2}\,{v_i}_{\,2n,2n+m}^{(k_1,k_2)}(X,Y)\,,
\\
\label{xiform}
x_i&=&\sum_{m,n=0}^{\infty}\hbar^{2n}\epsilon^{2n+m}
\sum_{k_1+k_2\atop=m}
\,V_x^{k_1}\,V_y^{k_2}\,{x_i}_{\,2n,2n+m}^{(k_1,k_2)}(X,Y)\,.
\end{eqnarray}
Again, the coefficient functions ${v_i}_{\,2n,2n+m}^{(k_1,k_2)}
(X,Y),{x_i}_{\,2n,2n+m}^{(k_1,k_2)}(X,Y)$ are gauge independent.

The products of symbols in
(\ref{Vgciform})--(\ref{xiform}) are defined point-wise.
To derive corresponding relations between the particle
operators $\hat{\bx},\hat{\bv}$
and the guiding center operators
$\hat{\bX},\hat{\bV}$ from (\ref{Vgciform})--(\ref{xiform}),
we have to introduce the star product on the right hand side
of these equations.
In (\ref{Vgciform}) and (\ref{Xgciform}) this is simply done
by making use of the correspondence rule
(\ref{Weylsymbgiex2}),(\ref{Weylsymbopgiex2}).
In the case of the backward transformations (\ref{viform}),
(\ref{xiform}) we have to ``translate'' the classical result
(\ref{xcl})--(\ref{vycl}) given in the appendix into operator
``language'' to obtain the lowest order term with respect to
$\hbar$.
It will turn out that this is enough to determine the lowest
order quantum mechanical correction to the guiding center
Hamiltonian.

To this end we first conclude from equation (\ref{Pcalformdef})
that for any two symbols $A_W$, $B_W$ their point-wise product
and star product differ by $\sum_{n=1}^{\infty}\hbar^n
{\cal P}_n(A_W,B_W)$.
Since the Moyal brackets of the components ($X,Y$) of the guiding
center position and the gyration velocity ($V_x,V_y$) vanish, we
have for two arbitrary functions $f(X,Y)$ and $g(V_x,V_y)$
\begin{equation}
\label{fgorder}
f(X,Y)*g(V_x,V_y) = g(V_x,V_y)*f(X,Y)
\end{equation}
and
\begin{equation}
\label{L2n1fg}
{\cal P}_{2n+1}\left[f(X,Y),g(V_x,V_y)\right] = 0\;,\;\;n
\in\Natural_0\,.
\end{equation} 
As an example, 
\begin{equation}
\label{sqrtBVx}
B^{1/2}(X,Y)*V_x = V_x*B^{1/2}(X,Y) = B^{1/2}(X,Y) V_x +
\hbar^2 {\cal P}_2 \!\left(B^{1/2},V_x\right) + {\cal O}
\left(\hbar^4\right),
\end{equation}
with ${\cal P}_2 \!\left(B^{1/2},V_x\right)$ being of order
$\epsilon$, so that the difference between $B^{1/2}V_x$ and
$B^{1/2}\!*\!V_x$ is of order $\hbar^2\epsilon$.
The fact that the symbols $x,y,v_x,v_y$ represent
self-adjoint operators leads directly to the
substitutions
\begin{equation}
\label{substitution}
V_xV_y \rightarrow \frac{1}{2}\left(V_x*V_y+V_y*V_x\right),
\;\;\;
V_x^2V_y \rightarrow V_x*V_y*V_x\;,
\;\;\;
V_y^2V_x \rightarrow V_y*V_x*V_y
\end{equation}
in equations (\ref{xcl})--(\ref{vycl}), because the symbols on
the right hand side of (\ref{substitution}) correspond to
self-adjoint operators.
Further computations show that the replacement of the
point-wise product with the star product in equations
(\ref{xcl})--(\ref{vycl}) leads to corrections which are at
least of order $\hbar^2\epsilon$ or $\hbar\epsilon^2$.
Therefore, up to terms of order $\epsilon^2$ the relations
between the particle operators (symbols) and the guiding center
operators (symbols) formally coincide with the classical result
(\ref{xcl})--(\ref{vycl}), if the point-wise product is replaced
by the star product and the substitution rules (\ref{substitution})
for products of gyration velocities are taken into account.
For this reason we refrain from writing them down explicitly
and refer the interested reader to the appendix.

Next we insert the results for $v_x,v_y$ into the
Hamiltonian (\ref{hamopsymb}) and expand the potential
$\phi(x,y)$ into a Taylor series around the guiding center
position ($X,Y$) replacing again point-wise products with
star products.
Naively one would expect from our previous results that the
symbol Hamiltonian is of the form
\begin{equation}
\label{hamsymbexp}
H_W = H_{cl} + {\cal O}(\epsilon^2\hbar, \epsilon \hbar^2)\,,
\end{equation}
with $H_{cl}$ being formally equal (in the sense described
above) to the classical Hamiltonian function
(\ref{hamfktallg}).
However, as pointed out earlier, $H_W$ should depend on the
gyration velocity components $V_x,V_y$ only by means of the
magnetic moment of gyration, $J\!=\!V_x*V_x\!+\!V_y*V_y$, and
its powers.
When evaluating the products $v_x\!*\!v_x$ and $v_y\!*\!v_y$
we have to change the ordering of the symbols $V_x,V_y$
accordingly.
Since their Moyal bracket is of order $\hbar \epsilon^{-1}$,
additional terms compared to the classical Hamiltonian
occur.
A straightforward calculation shows that the lowest order
correction originates from the $\epsilon^2$-term in the
classical Hamiltonian function.
It is of order $\hbar^2$.
Leaving away from now on the multiplication symbol ``$*$'' 
for reasons of notational simplicity, the Weyl symbol of the
guiding center Hamiltonian finally turns out to be
\begin{eqnarray}
\label{hamsymbres}
H_W\,=& &\!\!\!\!\!\!\frac{1}{2}\,B\left(\,V_x^2 + V_y^2
\right) + \phi(X,Y) \nonumber \\
&+&\frac{\epsilon^2}{16 B^2}\,\left[\left(-3 B_{,x}^2
+ B B_{,xx} - 3 B_{,y}^2 + B B_{,yy}\right)\left(\,V_x^2 +
V_y^2 \right)^2 \right.
\nonumber \\
& & \hspace{0.55in} + \left. 4 \left(3 E_x B_{,x} -
B E_{x,x} + 3 E_y B_{,y} - B E_{y,y} \right)
\left(V_x^2 + V_y^2\right) 
- 8 \left(E_{x}^2+E_{y}^2\right)\right] 
\nonumber\\
&+&\frac{\hbar^2}{16 B^2}\,\left(- B_{,x}^2 + B B_{,xx} -
B_{,y}^2 + B B_{,yy}\right)
+{\cal O}(\epsilon \hbar^2,\epsilon^2\hbar,\epsilon^3)\,.
\end{eqnarray}
Here, squares of symbols are star products of equal
factors, electric and magnetic fields have to be evaluated
at the guiding center position and the comma in a subscript
denotes differentiation with respect to the following
coordinate(s).
Using two-dimensional vector notation equation
(\ref{hamsymbres}) may be written in a more compact form,
\begin{eqnarray}
\label{hamsymbresvec}
H_W\,=& &\!\!\!\!\!\!\frac{1}{2}\,B J + \phi(X,Y)
 + \frac{\epsilon^2}{16 B^2}\,\left[\left( B \Delta B
-3\, | \nabla B |^2 \right)\! J^2 + 4 \left(
3\,\bE\!\cdot\!\nabla B - B\,\nabla\!\cdot\!\bE\right) J
- 8 \,|\bE|^2 \right] 
\nonumber\\
&+&\frac{\hbar^2}{16 B^2}\,\left( B \Delta B -
| \nabla B |^2 \right)
+{\cal O}(\epsilon \hbar^2,\epsilon^2\hbar,\epsilon^3)\,,
\end{eqnarray}
where scalar products denote summation over Weyl products
of vector components. 
Due to the use of the gauge invariant Weyl calculus, the
expansion of $H_W$ involves only gauge invariant quantities.
Thus, the higher order terms of the guiding center Hamiltonian
will be gauge independent as well. 

Up to second order in $\hbar$ and $\epsilon$ the
Hamiltonian $H_W$ in (\ref{hamsymbres}) depends on the gyration
velocities already via $V_x\!*\!V_x+V_y\!*\!V_y$, even
though we have not carried out the averaging transform
mentioned before.
The reason for this is that we used averaged classical
guiding center coordinates (as given in the appendix) as zero
order terms (with respect to $\hbar$) in the general expansion
(\ref{Vgciform}) of the guiding center symbols.

The term proportional to $\hbar^2$ in (\ref{hamsymbres})
specifies the lowest order quantum mechanical correction
to the classical guiding center Hamiltonian.
It does not depend on the electrostatic field
$\bE\!=\!-\nabla\phi$.
We will comment on its magnitude in the next section.

Employing the same procedure to the more general case of a
three-dimensional magnetic field and arbitrarily oriented
electric field, the leading quantum correction to the
classical Hamiltonian function turns out to be of second order
in $\hbar$ as well.
However, the results for the guiding center symbols and
the guiding center Hamiltonian are notationally cumbersome
and do not shed new light on our method.
Therefore we refrain from writing them down explicitly in
this paper.

\section*{4.\ Summary and conclusion}

\label{secIV}

There are two major results of our investigations: 
First the product rule for gauge invariant Weyl
symbols, equations (\ref{Lopdef}),(\ref{Lnopdef}), and
(\ref{Moyalrule1}).
They are a generalization of the well-known Moyal formula 
valid in the usual Weyl formalism.
The leading order term in the $\hbar$-expansion
(\ref{Moyalbrackexp}) of the Moyal bracket is proportional
to the Poisson bracket, which is expressed in terms of
derivatives with respect to position $\bx$ and kinetic
momentum $\bk$.
The higher order terms in (\ref{Moyalbrackexp}) are of
a more complex structure and cannot be written as powers
of the Poisson bracket operator.
The question about their interpretation in terms of
the modified phase space geometry in the presence of 
electromagnetic fields \cite{Ste77,Gui78,Wei78} may serve
as an interesting starting point for further, more
mathematical studies.

The Weyl symbol $H_W$ of the guiding center Hamiltonian,
equation (\ref{hamsymbres}), represents the second major
result of this paper.
The method used to derive it makes extensive use of the
product rule for gauge invariant Weyl symbols.
The great advantage of this approach lies in the fact that
the adiabatic parameter $\epsilon$ can be incorporated
into the gauge invariant Weyl calculus in a straightforward
manner.
Consequently, all expansions are carried out with respect
to both $\epsilon$ and Planck's constant $\hbar$.

Let us now investigate the importance of quantum mechanical
effects on guiding center motion by comparing the magnitudes
of the $\epsilon^2$- and $\hbar^2$-term in the guiding center
Hamiltonian (\ref{hamsymbres}) at low particle energies.
Taking into account the quantization of energy levels we
replace the gyration energies $\frac{1}{2}\,B( V_x^2\!+
\!V_y^2)$ with the corresponding harmonic oscillator
eigenvalues $(n\!+\!\frac{1}{2})\hbar \omega_B$, where
$\omega_B\!\equiv\!|qB|/(mc)$ is the cyclotron frequency
at the position of the guiding center.
In the absence of an electric field the guiding center
Hamiltonian for a spinless particle takes the form
(using vector notation)
\begin{equation}
\label{hamsymbquant}
H_W = (n + \frac{1}{2})\hbar\omega_B + 
\frac{\hbar^2(n+\frac{1}{2})^2}
{4 m B^2}\left( B \Delta B -3\, | \nabla B |^2 \right)
+ \frac{\hbar^2}{16 m B^2}\left( B \Delta B
- | \nabla B |^2 \right)\,,
\end{equation}
with correct physical units restored in the way explained at
the beginning of section 3.
The second term in (\ref{hamsymbquant}) is the adiabatic
correction (with $\epsilon$ set to unity) which now contains
$\hbar$ due to energy quantization.
Only for small oscillator (=gyration) quantum numbers $n$ the
lowest order classical and quantum mechanical corrections
are of the same magnitude, otherwise the classical term
dominates.
From classical guiding center theory, however, it is known that
at low particle energies the influence of adiabatic corrections
on guiding center motion is negligibly small.
Since the magnitude of the leading quantum mechanical correction
in (\ref{hamsymbquant}) does not depend on the particle energy,
we conclude that quantum effects on guiding center motion may be
neglected at all energy scales.
Thus, when carrying out experiments with charged particles in
inhomogeneous magnetic fields even at very low temperatures, the
motion of the guiding center is described excellently by the
lowest order classical equations.
This is also true if an additional electrostatic field $\bE$
is switched on, because according to  (\ref{hamsymbres}) the
magnitude of the lowest order quantum mechanical correction does
not depend on $\bE$.
As a general result of our investigations we may therefore
say that guiding center motion is not effected by quantum
mechanics.

Note that if the quantum number $n$ becomes too large in
equation (\ref{hamsymbquant}), the adiabatic correction
dominates over the first (gyrative) term in the Hamiltonian.
This parallels the breakdown of classical guiding center
theory at large particle energies. 

For reasons of completeness, let us finally compare our result
(\ref{hamsymbres}) for the guiding center Hamiltonian with the
quantum mechanical calculation \cite{Mar96b} of Maraner, who
investigated the motion of a charged spinning particle in a
two-dimensional magnetic field.
The interaction between the magnetic moment $\bmu$ of the
particle and the external magnetic field is included into
the Hamiltonian by the potential
$-\bmu\!\cdot\!\bB$.
Since the component $\mu_z$ of the magnetic moment
parallel to $\bB\!=\!
B(x,y)\be_z$
is a constant of motion we may replace
$-\bmu\!\cdot\!\bB$
by the scalar term $-\mu_z B(x,y)$.
Thus, the classical Hamiltonian function reads
\begin{equation}
\label{hamopspin}
H =\frac{1}{2}\left(\,v_x^2 + v_y^2\right)
-\mu_z B(x,y)\,.
\end{equation}
The term $-\mu_z B(x,y)$ may be interpreted as a special
case of a time-independent scalar potential $\phi(x,y)$.
A straightforward calculation shows that the corresponding
guiding center symbol Hamiltonian is given by
\begin{eqnarray}
\label{HamsymbmuB}
H_W'\,=& &\!\!\!\!\!\!\frac{1}{2}\,B J - \mu_z B
+ \frac{\epsilon^2}{16 B^2}\,\left[
\left( B \Delta B -3\, | \nabla B |^2 \right)
J\left(\,J - 4 \mu_z\right)
- 8\,\mu_z^2 B^{-2} | \nabla B |^2 \right] 
\nonumber\\
&+&\frac{\hbar^2}{16 B^2}\,
\left( B \Delta B - | \nabla B |^2 \right)
+{\cal O}(\epsilon \hbar^2,\epsilon^2\hbar,\epsilon^3)\,,
\end{eqnarray}
where the magnetic field and its derivatives have to be
taken at the guiding center position.
Substituting $\mu_z\!=\!-g\hbar\sigma_3$ ($g$ =
gyromagnetic factor of the particle, $\sigma_3\!=\!\pm1$
for spin-$\frac{1}{2}$ particles),
$H_W'$ takes the same form as the Hamiltonian operator
derived by Maraner.
However, in (\ref{HamsymbmuB}) there are two expansion
parameters $\epsilon$ and $\hbar$ distinguishing between
adiabatic and quantum mechanical corrections to the guiding
center Hamiltonian whereas in \cite{Mar96b} the only
expansion parameter is the magnetic length $l_B=\sqrt{\hbar
c/(q |\bB|)}$.

\section*{Acknowledgments}
\label{acknowledge}
The author wishes to thank Robert Littlejohn for his kind
hospitality at the Lawrence Berkeley Laboratory and many
helpful suggestions.
Thanks also to Jim Morehead and Kevin Mitchell for several
interesting discussions about the Wigner-Weyl formalism and
carefully reading this manuscript. 
This work was performed under the financial support of the
Deutsche Forschungsgemeinschaft (DFG) which is gratefully
acknowledged.

\appendix
\section*{}

In this appendix we will briefly summarize the classical
guiding center theory for the motion of a charged particle
in a two-dimensional magnetic field $\bB\!=\!
B(x,y)\be_z$ and an electrostatic field
$\bE\!=\!-\nabla\phi(x,y)$.
We closely follow the Hamiltonian method developed by
Littlejohn which employs non-canonical, but gauge invariant
coordinates in phase space.
For further details the interested reader is referred to
\cite{Lit79,Lit81a}.

To incorporate the classical result into the
symbol calculus we have to determine the Cartesian components
$V_x,V_y$ of the gyration velocity instead of the
generalized gyrophase $\theta$ and magnetic momentum $J$.
The reason lies in the difficulty of defining a quantum
mechanical operator corresponding to the classical angle
variable $\theta$.
This choice of guiding center phase space coordinates has
the disadvantage that in order to compute ($X,Y,V_x,V_y$)
we cannot apply the elegant geometric method of
\cite{Lit79,Lit81a}.

Introducing the adiabatic parameter $\epsilon$ in the
standard way, i.e.\ by replacing the charge $q$ by
$q/\epsilon$, and scaling physical quantities according to
equation (\ref{physcal}) we are looking for phase space
functions ($X,Y,V_x,V_y$) whose non-vanishing Poisson
brackets are given by
\begin{equation}
\label{gcPb}
\{V_x,V_y\} = \frac{1}{\epsilon}\;,\;\;
\{X,Y\} = \frac{\epsilon}{B(X,Y)}\,.
\end{equation}
Again, their different order with respect to $\epsilon$ 
indicates the different time scales of motion.
Making the perturbative ansatz
\begin{equation}
\label{gccoordpert}
Z_i = \sum_{k=0}^{\infty}\epsilon^k Z_i^k(x,y,v_x,v_y)
\end{equation}
for the guiding center phase space coordinates
$\left\{Z_i\right\}_{1\leq i \leq 4}\!=\!\left(X,Y,V_x,V_y
\right)$ we may calculate the coefficient functions
$Z_i^k(x,y,v_x,v_y)$ order by order by solving the partial
differential equations implied by the Poisson bracket
relations (\ref{gcPb}).
To make sure that the functions $Z_i^k(x,y,v_x,v_y)$ do not
depend on the gauge, we have to write down the Poisson
bracket in a gauge invariant form,
\begin{equation}
\label{Pbbxbv}
\left\{f,g\right\} = \frac{\partial f}
{\partial \bx}
\cdot \frac{\partial g}{\partial \bv}\,-\,
\frac{\partial f}{\partial \bv}
\cdot \frac{\partial g}{\partial \bx}
\,+\, \frac{1}{\epsilon}\,\bB\!\cdot\!
\left(\frac{\partial f}{\partial \bv}
\times \frac{\partial g}{\partial \bv}
\right),
\end{equation}
using only derivatives with respect to the physical (i.e.\
gauge independent) phase space coordinates of the particle,
namely its position $\bx$ and velocity $\bv$.

The resulting guiding center coordinates are not uniquely
determined by the Poisson bracket relations.
In order to facilitate the following computations it is
of advantage to choose them in the simplest form possible.
To lowest order they are proportional to the related particle
coordinates.
Therefore, $V_x,V_y$ are rapidly oscillating functions of time
because the particle velocity components $v_x,v_y$ depend on
the gyration angle $\theta$.

Next we have to find the corresponding backward
transformations, i.e.\ to write the particle coordinates
($x,y,v_x,v_y$) as functions of ($X,Y,V_x,V_y$).
Inserting the result into the Hamiltonian function
\begin{equation}
\label{hamfktcl}
H =\frac{1}{2}\left(\,v_x^2 + v_y^2\right) + \phi(x,y)\,,
\end{equation}
the latter takes the form of an asymptotic series with
respect to $\epsilon$ with the coefficients being gauge
invariant functions of the guiding center coordinates.
To exclude rapidly oscillating terms in this expansion
the Hamiltonian should depend on the gyration velocity
components only by means of $J\!=\!V_x^2\!+\!V_y^2$ and
its powers, because to lowest order in $\epsilon$ $J$ does
not depend on the gyration angle $\theta$. 
This can be achieved by carrying out a near-identity coordinate
transformation $(X,Y,V_x,V_y)\rightarrow(\bar X,\bar Y,
\bar V_x,\bar V_y)$ to a new set of (averaged) guiding
center coordinates. 
They have to satisfy the same Poisson bracket relations
(\ref{gcPb}) as the old ones so that the related magnetic
moment $\bar J\!=\!\bar V_x^2\!+\!\bar V_y^2$ becomes a
constant of motion and at the same time the dynamics of the
guiding center position decouples from that of the gyration.
Such a kind of symplectic transformation may be expressed in
terms of Lie generators (cf.\ \cite{Lit79,Lit81a}).

As a result, up to second order in $\epsilon$ the averaged
guiding center coordinates read (leaving away the bar over
them and using commas in the subscripts to denote
differentiation with respect to the following coordinate(s))
\begin{eqnarray}
\label{Xcl}
X\,=& &\!\!\!\!\!\!x\,+\,\epsilon\,B^{-1}v_y\,+\,
\frac{\epsilon^2}{2B^3}\left(B_{,y} v_x - B_{,x} 
v_y \right) v_y\,+\,{\cal O}(\epsilon^3),
\\
\label{Ycl}
Y\,=& &\!\!\!\!\!\!y\,-\,\epsilon\,B^{-1}v_x\,-\,
\frac{\epsilon^2}{2B^3}\left(B_{,y} v_x - B_{,x}
v_y \right) v_x\,+\,{\cal O}(\epsilon^3),
\\
\label{Vxcl}
V_x\,=& &\!\!\!\!\!\!B^{-1/2}\,v_x\,+\,
\frac{\epsilon}{2 B^{5/2}}\,\left(B_{,y} v_x^2 +
B_{,x} v_x v_y + 2 B_{,y} v_y^2 + B B_{,y} 
\right)
\nonumber \\ 
&+&\,\frac{\epsilon^2}{16 B^{9/2}}\left[\!
\frac{}{}\left(-5B_{,x}^2 - B B_{,xx} + 13 B_{,y}^2
- 5 B B_{,yy}\right) v_x^3 - 2\left(4 B_{,x} B_{,y}
+ 2 B B_{,xy} - c_1\right) v_x^2 v_y\right.
\nonumber \\
&+& \left.\left(-15 B_{,x}^2 
+ 7 B B_{,xx} + 23 B_{,y}^2 - 13 B B_{,yy} \right)
v_x v_y^2 + 2 \left(- 14 B_{,x} B_{,y} + 6 B B_{,xy} 
+ c_1\right)v_y^3
\right. \nonumber \\
&+&  \left.4 B \left(B_{,x}E_x 
+ B E_{x,x} - 7 B_{,y} E_y - 3 B E_{y,y}\right) v_x  
+ 2 B \left( c_2 + 9 B_{,x} E_{y} + 9 B_{,y} E_{x}
\right.\right.
\nonumber \\
&-&\left.\left. 8 B E_{x,y} \right) v_y \,\frac{}{}\right]
+ {\cal O}(\epsilon^3)\,, \\
\label{Vycl}
V_y\,=& &\!\!\!\!\!\!B^{-1/2}\,v_y\,-\,
\frac{\epsilon}{2 B^{5/2}}\,\left(2 B_{,x} v_x^2 
+ B_{,y} v_x v_y + B_{,x} v_y^2 + B B_{,x} \right)
\nonumber \\ 
&+&\,\frac{\epsilon^2}{16 B^{9/2}}\!\left[\frac{}{}
2 \left(- 14 B_{,x} B_{,y} + 6 B B_{,xy} 
- c_1\right) v_x^3 +\left(23 B_{,x}^2 -13 B B_{,xx} 
- 15 B_{,y}^2  + 7 B B_{,yy} \right)\!\times\right. 
\nonumber \\
& & v_x^2 v_y - \left.2\left( 4 B_{,x} B_{,y} 
+ 2 B B_{,xy}+ c_1 \right) v_x v_y^2 + \left( 13 B_{,x}^2
- 5 B B_{,xx} - 5 B_{,y}^2 -  B B_{,yy}\right) v_y^3
\right.
\nonumber \\
&-&  \left. 2 B \left( c_2  
- 7 B_{,x} E_{y} - 7 B_{,y} E_{x} \right) v_x 
+ 4 B \left( - 7 B_{,x} E_x + 3 B E_{x,x} 
+ B_{,y} E_y + B E_{y,y}
\right) v_y \,\frac{}{}\right]
\nonumber \\
&+&{\cal O}(\epsilon^3),
\end{eqnarray}
with the corresponding backward transformation
\begin{eqnarray}
\label{xcl}
x\,=& &\!\!\!\!\!\!X\,-\,\epsilon\,B^{-1/2}V_y\,-\,
\frac{\epsilon^2}{2B^2}
\left(2 B_{,x} V_x^2 + B_{,y} V_x V_y + B_{,x} V_y^2 
- 2 B_{x} \right) \,+\,{\cal O}(\epsilon^3),
\\
\label{ycl}
y\,=& &\!\!\!\!\!\!Y\,+\,\epsilon\,B^{-1/2}V_x\,-\,
\frac{\epsilon^2}{2B^2}
\left( B_{,y} V_x^2 + B_{,x} V_x V_y + 2 B_{,y} V_y^2
- 2 E_{y} \right)\,+\,{\cal O}(\epsilon^3),
\\
\label{vxcl}
v_x\,=& &\!\!\!\!\!\!B^{1/2}\,V_x\,-\,
\frac{\epsilon}{B}\left[ \left( B_{,x} V_x 
+ B_{,y} V_y \right) V_y - E_{y} \right]
\nonumber \\
&+&\,\frac{\epsilon^2}{16 B^{5/2}} \left[\!\frac{}{}
\left(- 11 B_{,x}^2 + B B_{,xx} - 3 B_{,y}^2 
+ B B_{,yy}\right) V_x^3 - 4\left(5 B_{,x} B_{,y} 
+ B B_{,xy} + c_1/2\right)\times\right. 
\nonumber \\
& & V_x^2 V_y + \left. \left( B_{,x}^2 
+ 5 B B_{,xx} - 15 B_{,y}^2
- 3 B B_{,yy} \right) V_x V_y^2 + 4 \left( B_{,x} B_{,y}
+ B B_{,xy} - c_1/2 \right) V_y^3
\right. \nonumber \\
&+&  \left. 4 \left( 3 B_{,x} E_x - B E_{x,x} 
+ B_{,y} E_y + B E_{y,y}\right) V_x 
- 2 \left( c_2 - B_{,x} E_{y} - B_{,y} E_{x} \right)
V_y\,\frac{}{}\right] \nonumber \\
&+&{\cal O}(\epsilon^3), \\
\label{vycl} 
v_y\,=& &\!\!\!\!\!\!B^{1/2}\,V_y\,+\,
\frac{\epsilon}{B}\,\left[ \left( B_{,x} V_x 
+ B_{,y} V_y \right) V_x - E_{x} \right]
\nonumber \\ 
&+&\,\frac{\epsilon^2}{16 B^{5/2}} \left[\frac{}{}
4 \left( B_{,x} B_{,y} + B B_{,xy} + c_1/2\right) V_x^3 
+ \left(-15 B_{,x}^2 - 3 B B_{,xx}
+ B_{,y}^2 + 5 B B_{,yy} \right)\times \right.
\nonumber \\
& &  V_x^2 V_y + \left. 4\left( 5 B_{,x} B_{,y} 
+  B B_{,xy} - c_1/2 \right) V_x V_y^2 
+ \left( - 3 B_{,x}^2 + B B_{,xx}  - 11 B_{,y}^2 
+ B B_{,yy}\right) V_y^3 
\right. \nonumber \\
&+& \left.2 \left( c_2  + 3 B_{,x} E_{y} 
+ 3 B_{,y} E_{x} - 8 B E_{x,y}\right)
V_x + 4 \left( B_{,x} E_x + B E_{x,x} + 3 B_{,y}
E_y - B E_{y,y} \right) V_y \,\frac{}{}\right] 
\nonumber \\
&+&{\cal O}(\epsilon^3).
\end{eqnarray}
In (\ref{Vxcl}), (\ref{Vycl}), (\ref{vxcl}), (\ref{vycl}),
$c_1$ and $c_2$ are arbitrary constants which remain
uneffected by the requirement that the Hamiltonian has
to be independent of the gyration angle up to second
order in $\epsilon$ (for more details concerning the
ambiguity of guiding center coordinates cf.\ 
\cite{Lit82p}).
In (\ref{xcl})--(\ref{vycl}) the fields and their
derivatives have to be evaluated at the guiding center
position ($X,Y$).

In terms of the guiding center coordinates, the Hamiltonian
reads 
\begin{eqnarray}
\label{hamfktallg}
H\,=& &\!\!\!\!\!\!\frac{1}{2}\,B\left(\,V_x^2 + V_y^2 
\right) + \phi(X,Y) \nonumber \\
&+&\frac{\epsilon^2}{16 B^2}\,\left[\left(-3 B_{,x}^2
+ B B_{,xx} - 3 B_{,y}^2 + B B_{,yy}\right)\left(\,V_x^2
+ V_y^2 \right)^2 \right.\nonumber \\
& & \hspace{0.55in} + \left. 4 \left(3 E_x B_{,x} - 
B E_{x,x} + 3 E_y B_{,y} - B E_{y,y} \right)
\left(V_x^2 + V_y^2\right) 
- 8 \left(E_{x}^2+E_{y}^2\right)\right] \nonumber\\
&+&{\cal O}(\epsilon^3)\,,
\end{eqnarray}
which may be written in a more compact form by employing
two-dimensional vector notation, 
\begin{eqnarray}
\label{hamfktallgvec}
H\,=& &\!\!\!\!\!\!\frac{1}{2}\,B J + \phi(X,Y)
 + \frac{\epsilon^2}{16 B^2}\,\left[\left( B \Delta B
-3\, | \nabla B |^2 \right) J^2 + 4 \left(
3\,\bE\!\cdot\!\nabla B - B\,\nabla\!\cdot\!\bE\right) J
- 8 \,|\bE|^2 \right] \nonumber\\
&+&{\cal O}(\epsilon^3)\,.
\end{eqnarray}
In (\ref{hamfktallg}) and (\ref{hamfktallgvec}) the fields
and potentials have to be evaluated at the guiding center
position ($X,Y$).
For obvious reasons the last two expressions are denoted as
the classical guiding center Hamiltonian.

\end{document}